\documentclass[usenatbib,useAMS]{mnras}

\pdfoutput=1

\usepackage[T1]{fontenc}
\usepackage{ae,aecompl}

\usepackage{graphicx}	
\usepackage{amsmath}	
\usepackage{amssymb}	

\newcommand{\lya}{Ly $\alpha$}

\newcommand{\dtbbar}{\ensuremath{\overline{\delta T_\mathrm{b}}}}

\title[Parametrizations of the global 21-cm signal]{Parametrizations of the 21-cm global signal and parameter estimation from single-dipole experiments}

\author[Harker et al.]{Geraint J.A. Harker$^{1,4}$\thanks{E-mail: g.harker@ucl.ac.uk}, Jordan Mirocha$^{2,5}$, Jack O. Burns$^{2,6}$,\newauthor Jonathan R.~Pritchard$^3$\\$^1$Department of Physics and Astronomy, University College London, London WC1E 6BT, UK \\$^2$Center for Astrophysics and Space Astronomy, Department of Astrophysical and Planetary Sciences,\\$^{\phantom{2}}$University of Colorado, Campus Box 389, Boulder, CO 80309\\$^3$Department of Physics, Blackett Laboratory, Imperial College, London SW7 2AZ, UK\\$^{4}$Marie Curie Fellow\\$^{5}$NASA Earth and Space Sciences Graduate Fellow\\$^{6}$NASA Ames Research Center, Moffett Field, CA 94035}

\date{Accepted XXX. Received YYY; in original form ZZZ}

\pubyear{2015}

\begin{document}
\label{firstpage}
\pagerange{\pageref{firstpage}--\pageref{lastpage}}
\maketitle

\begin{abstract}
One approach to extracting the global 21-cm signal from total-power measurements at low radio frequencies is to parametrize the different contributions to the data and then fit for these parameters. We examine parametrizations of the 21-cm signal itself, and propose one based on modelling the \lya\ background, intergalactic medium temperature and hydrogen ionized fraction using tanh functions. This captures the shape of the signal from a physical modelling code better than an earlier parametrization based on interpolating between maxima and minima of the signal, and imposes a greater level of physical plausibility. This allows less biased constraints on the turning points of the signal, even though these are not explicitly fit for. Biases can also be alleviated by discarding information which is less robustly described by the parametrization, for example by ignoring detailed shape information coming from the covariances between turning points or from the high-frequency parts of the signal, or by marginalizing over the high-frequency parts of the signal by fitting a more complex foreground model. The fits are sufficiently accurate to be usable for experiments gathering $1000\ \mathrm{h}$ of data, though in this case it may be important to choose observing windows which do not include the brightest areas of the foregrounds. Our assumption of pointed, single-antenna observations and very broad-band fitting makes these results particularly applicable to experiments such as the {\it Dark Ages Radio Explorer}, which would study the global 21-cm signal from the clean environment of a low lunar orbit, taking data from the far side.
\end{abstract}




\section{Introduction} 
\label{sec:intro}

The sky-averaged or `global' 21-cm signal, $\dtbbar(z)$, is the mean differential brightness temperature of the 21-cm line of hydrogen, relative to the cosmic microwave background (CMB), as a function of redshift or observed frequency. The amplitude of the signal is determined by the amount of neutral hydrogen present and by the relative number of electrons in the ground and excited states of the 21-cm transition, which is often defined in terms of the spin temperature, $T_\mathrm{s}$, of the transition. $T_\mathrm{s}$ depends in turn on the kinetic temperature, $T_\mathrm{k}$, of the gas, and on the efficiency with which $T_\mathrm{s}$ can be driven towards $T_\mathrm{k}$, and away from the CMB temperature, by collisions and by \lya\ coupling \citep[the Wouthuysen-Field effect;][]{wouthuysen1952,field1958}. $\dtbbar(z)$ is, therefore, sensitive to the evolution of a variety of different radiation fields: ionizing radiation, which destroys neutral hydrogen; X-rays, which can heat the gas and raise $T_\mathrm{k}$; and \lya, which causes Wouthuysen-Field coupling. This makes it a valuable probe of sources of radiation and heating up to the end of the epoch of reionization at $z \approx 6$ \citep[see e.g.,][and references therein]{pritchard2012}.

To make inferences about the properties of the high-redshift Universe from radio observations at the relevant frequencies ($<200\ \mathrm{MHz}$), it is useful to have models for $\dtbbar(z)$ that can be fit to the data and compared with each other. We note that some of the information contained in the signal, especially about the radiation fields which affect it most directly, may be extracted in a relatively model-independent way \citep{mirocha2013}, and that it may be possible to extract the signal as a residual without explicitly fitting for it during foreground removal \citep{liu2013,switzer2014}. Nonetheless, we generally deal with parametrized models. These parameters may be physical --- describing assumptions about, for example, the spectra of early sources, the efficiency of different types of star formation and the mass of star-forming haloes --- and we then require a code which can compute the radiation backgrounds from the evolving population of sources, and their effect on the 21-cm signal. At the other end of the scale, we can choose a flexible parametric form which we hope can describe the shape of $\dtbbar(z)$ without having to specify the details of the physics in advance, such as a cubic spline, as in \citet{pritchard2010}.

The fitting is complicated by the fact that the redshifted 21-cm spectrum is superimposed on that of bright astrophysical foregrounds, and is observed by an instrument with a frequency response which may be complicated and/or known only through careful calibration.  One approach to this problem is to find parametrized models for the foregrounds, instrument, and any other components of the observed spectrum (or spectra, if multiple pointings are used in order to gather more independent information on the foregrounds and the instrument), and fit them simultaneously with the parameters of the 21-cm signal. This entails searching what may be a high-dimensional parameter space, which one may do with a Markov Chain Monte Carlo sampler \citep{harker2012}, nested sampling \citep{harker2015} or similar methods. This allows us to rigorously characterize the errors, study the degeneracies between foreground and signal parameters, and so on, but may require the likelihood (and therefore a realization of the signal) to be computed many times. This places requirements on the computational cost of our signal model, as well as on its flexibility and accuracy.

The aim of this paper is to study different parametrizations of $\dtbbar(z)$, comparing their ability to represent the signal faithfully, to retain the astrophysical information in the measured spectrum, and to be used in a signal extraction pipeline where the cost of computing the model signal may be important. This is timely because a number of projects (current or proposed) aim to detect the 21-cm global signal, e.g.\ EDGES \citep{rogers2015}, LEDA \citep{greenhill2012}, BIGHORNS \citep{sokolowski2015} and SCI-HI \citep{voytek2014}. EDGES, especially, has already produced tentative constraints on the rapidity of reionization ($\Delta z>0.06$) from observations between 100 and 200 MHz \citep{bowman2010}. We describe the setup of these experiments in general terms in Section~\ref{sec:dareprob}, then, for concreteness, focus more specifically on the {\it Dark Ages Radio Explorer} ({\it DARE}; \citealt{burns2012}; Burns et al. in preparation). Our parametrizations of the signal are introduced in Section~\ref{sec:paramns}, and we then describe the process of extracting the signal from the data in Section~\ref{sec:description}. The quality of the recovery is compared for different parametrizations and experimental setups in Section~\ref{sec:recovery}, which we discuss further in Section~\ref{sec:disc} before offering some conclusions in Section~\ref{sec:conc}.

\section{Experimental setup} 
\label{sec:dareprob}

We consider a pointed experiment, that is it takes integrated low-frequency radio spectra in a number of discrete directions rather than, say, adopting some sort of scanning approach. We therefore concentrate on the case where we have some small number of independent spectra from which we wish to extract the global signal. In the case of {\it DARE}, the maximum number of independent pointings is $\sim 8$, since the antenna power pattern is broad (the beam has a full width at half-maximum of tens of degrees, depending on frequency). All else being equal, a larger number of pointings should improve our constraints, since they provide more independent information on the foregrounds and the instrument. If a very large fraction of the sky is covered, however, this implies that some spectra will include brighter regions near the Galactic Centre, reducing our sensitivity. This implies a tradeoff which we examine later.

In designing a global signal experiment, including designing a suitable antenna, the optimal frequency range must also be considered. A very low frequency experiment to study the dark ages would encounter significant problems if conducted from the ground, because of the ionosphere \citep{datta2014,vedantham2014}, but might also require a large antenna which could cause problems for a space mission, so we do not consider it here. We wish to include the start of the cosmic dawn in our analysis, which suggests starting at $40\ \mathrm{MHz}$ or lower, while a practical antenna can offer sensitivity and a smooth beam over perhaps a factor of 3 in frequency, suggesting a maximum of around $120\ \mathrm{MHz}$. This will probably allow an experiment to capture the start of reionization, but not the end. We will generally consider a frequency window of $35$--$120\ \mathrm{MHz}$, though we will also look at the effects of narrower ranges.

At each frequency, the sky temperature seen by the antenna is an integral of the true sky over the antenna power pattern. The true sky includes both the 21-cm signal and the foreground signal. The parametrization of the signal is our main concern here, and so we treat the foregrounds very simply, assuming they can be described by a polynomial in $\log(\nu)$--$\log(T)$ for each pointing, where $\nu$ is frequency. The coefficients of this polynomial, which constitute the parameters of our foreground model, are computed from the global sky model of \citet{dOC2008}. This is done by first integrating the sky model over the power pattern of the antenna at a large number of frequencies for each pointing we wish to consider, and then computing a polynomial fit in $\log(\nu)$--$\log(T)$. Because of this subsequent fitting stage, the power pattern we assume for the antenna is not important; we take it to be Gaussian, so that the computation can be done efficiently using the routines in \textsc{healpy}\footnote{\url{https://github.com/healpy/healpy}}, which is based on the \textsc{healpix} \citep{gorski2005} package\footnote{\url{http://healpix.sourceforge.net/}}.

We assume a simple model for the instrument response: the calibrated noise-free (modelled) spectrum is given by
\begin{equation}
  T_\mathrm{mod}(\nu) = G(\nu)T_\mathrm{sky}(\nu)+T_\mathrm{rcv}(\nu) ,
  \label{eqn:instresp}
\end{equation}
where $T_\mathrm{rcv}$ is the receiver temperature, which we assume to be a constant. The noise on a frequency channel of width $\Delta\nu$ after an integration time $t_\mathrm{obs}$ is then given by the radiometer equation, $\sigma=T_\mathrm{mod}/\sqrt{2t_\mathrm{obs}\Delta\nu}$, assuming two polarizations are averaged together. For the high sky temperatures at these frequencies, $T_\mathrm{rcv}$ therefore makes a relatively minor contribution to the noise temperature unless $G$ is very small, so although we assume it to be $100\ \mathrm{K}$ for concreteness, this has little influence on our results.



\section{The 21-cm global signal and parametrizations} 
\label{sec:paramns}

Our model for the 21-cm signal is computed with the Accelerated Reionization Era Simulations (\textsc{ares}) code\footnote{\url{https://bitbucket.org/mirochaj/ares}}, first designed to investigate the signatures of X-ray heating in the global 21-cm signal \citep{mirocha2014}. As in previous works \citep[e.g.,][]{furlanetto2006,pritchard2010}, \textsc{ares} divides the intergalactic medium (IGM) into two phases: (i) a fully ionized phase representing \ion{H}{ii} bubbles around galaxies, whose volume filling factor $Q_{\ion{H}{ii}}$ affects the overall normalization of the global 21-cm signal, and (ii) a mostly neutral `bulk IGM' phase beyond bubbles, whose spin temperature determines the strength and sign of the global 21-cm signal.

There are many parameters in \textsc{ares} that can be varied to generate different realizations of the 21-cm signal. In this work, we consider the cosmological parameters and the primordial power spectrum to be fixed, which in principle specifies the amount of matter which has been confined in haloes (and the mass function of haloes) as a function of redshift. The remaining parameters of the model govern how efficiently this collapsed mass is converted into radiation of different wavelengths, including the \lya\ which causes Wouthuysen-Field coupling, ionizing radiation (which affects $Q_{\ion{H}{ii}}$), and X-rays which heat the IGM efficiently. For example, we may specify a minimum halo virial temperature below which a halo cannot form stars, the formation efficiency and spectral energy distribution of different source populations, the fraction of the radiation which escapes into the IGM, and so on. Our reference model assumes values for these parameters which are consistent with low-redshift values, and results in a history for which the Thomson scattering optical depth to the CMB is consistent with constraints from \textit{Planck} \citep{planck2015}.

Part of the aim of this series of papers is to determine how well the values of these parameters really can be expected to represent the physics generating the signal. That is, if we generate a signal with \textsc{ares}, and use it as part of a synthetic data set, to what extent can we recover what we put in? This includes recovery of (i) the signal itself, (ii) the properties of the IGM consistent with that recovered signal, and ultimately, (iii) the properties of galaxies required to explain the IGM properties.

In its simplest mode of operation -- in which the cosmological radiative transfer is treated approximately -- \textsc{ares} is in principle fast enough to be used to model the signal when fitting a synthetic data set, which simultaneously yields a recovered signal, the entire history of \lya\ emission, ionization and heating in the IGM, and constraints on the values for physical parameters in the code. To fit more computationally expensive physical models, which have more free parameters, more advanced physics, or both, may require a ``two-stage'' approach, in which a preliminary fit is performed using a computationally efficient -- but perhaps phenomenological -- parametrization of the global 21-cm signal. This first stage yields a set of measurements to be fit subsequently by a more complex model. We will now consider two parametrizations of the signal that are inexpensive alternatives to the full \textsc{ares} model, and revisit the two-stage approach in \S\ref{sec:recovery}.

The first parametrization that we will consider is that put forward by \citet{pritchard2010}, which we will call the `turning points' parametrization. This has also been used in subsequent studies \citep{harker2012,harker2015} as a fast and convenient method to describe the major features of a generic global signal. In this picture, the 21-cm spectrum has a number of maxima and minima, caused as different effects become dominant in determining \dtbbar. The positions (in redshift or frequency, and in brightness temperature) of these turning points are the parameters of the model. The signal between these turning points is modelled as a cubic spline. This parametrization is very flexible and can describe a wide range of plausible 21-cm signals, but the turning point positions require further interpretation in order to relate them to the physics of the first sources, via their constraints on the intensity of various radiation fields and the properties of the IGM at the redshifts of the turning points \citep{mirocha2013}.

\begin{table*}
 \caption{The turning points parametrization models a signal with five maxima and minima, which are described here, along with their positions in the history produced by the reference \textsc{ares} model. When fitting using the turning points parametrization, the positions of turning points A and E are fixed, while the frequency and amplitude of turning points B, C and D are parameters.}
 \label{tab:TP_desc}
 \begin{tabular}{@{\quad}cccccl}
   \hline
   Label & $\nu/\mathrm{MHz}$ & $z$ & $\delta T_\mathrm{b}/\mathrm{mK}$ & Type & Description \\
   \hline
   A & 16.1 & 87.2 & $-42$ & Minimum & `Dark ages'; collisional coupling becomes ineffective.\\
   B & 47.4 & 29.0 & $-4.4$ & Maximum & `Cosmic dawn'; \lya\ coupling becomes effective. \\
   C & 71.0 & 19.0 & $-125$ & Minimum & Heating becomes important. \\
   D & 111.4 & 11.7 & 19.2 & Maximum & Heating saturated; reionization begins. \\
   E & 180 & 6.9 & 0 & Minimum & End of reionization; null signal after this point. \\
   \hline
 \end{tabular}
\end{table*}

We will also consider a parametrization which is in some sense intermediate between the `turning points' model and physical models like \textsc{ares},  which we will call the `tanh' parametrization. In this model, the \lya\ background (which determines the strength of Wouthuysen-Field coupling), the temperature of the IGM, and the ionized fraction of hydrogen, all evolve according to a tanh model, i.e.\ for each quantity $A(z)$ we have
\begin{equation}
  A(z)=\frac{A_\mathrm{ref}}{2}\big\{1+\tanh[(z_0-z)/\Delta z]\big\} ,
  \label{eqn:tanhdef}
\end{equation}
  so that they are zero at high redshift\footnote{With the exception of the IGM temperature, which is a sum of a `tanh' term and an adiabatic cooling term, $T_\mathrm{K} \propto (1 + z)^2$, in order to reproduce the `dark ages' signal prior to first light.}, switch on over an interval of width $\Delta z$ around a redshift of $z_0$, and become saturated at a value of $A_\mathrm{ref}$ at low redshift. Therefore each of these three quantities has three parameters describing its evolution, though e.g.\ the  $A_\mathrm{ref}$ parameter for ionization has a natural value of unity and is not free to vary if the model is to represent a physical history. Because the tanh model specifies values for IGM quantities, unlike the `turning points' model which is purely phenomenological, it can also yield e.g.\ the Thomson optical depth. The parameters are not linked directly to source properties, however, and so it is in this sense that we describe it as being intermediate between the \textsc{ares}-based parametrization and the `turning points' parametrization.

\begin{figure}
\includegraphics[width=\columnwidth,clip=true]{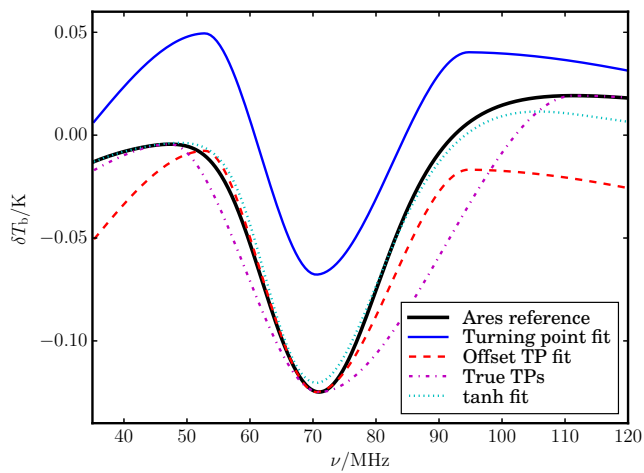}
    \caption{The ability of different parametrizations to fit the \textsc{ares} model is compared. The solid, black line shows the input model used to generate the synthetic data set, which also assumes  foregrounds modelled as third-order polynomials in log-frequency--log-temperature, an idealized instrument model in which the antenna has uniform 85 per cent sensitivity between 35 and 120~MHz, and an experiment which observes four disjoint sky regions for 250~h each. The recovered signal using the turning points parametrization is shown is the solid blue line; if we shift this down so that the temperature of turning point C agrees with the input signal, we have the dashed red line. If we use the actual positions of the maxima and minima of the \textsc{ares} signal as the parameter values in our turning points model, we produce the magenta dot--dashed curve. Finally, if the synthetic data set is fit using the tanh model, the signal we recover is shown as the cyan, dotted curve.}\label{fig:comparison}
\end{figure}

Although we will introduce our fitting procedure fully in the next section, we show the result of attempting to fit the signal generated by \textsc{ares} with the turning points and tanh parametrizations now, in order to point out some features of the parametrizations. This fitting, shown in Fig.~\ref{fig:comparison}, assumes an idealized instrument model, simple foregrounds, and 1000~h of observation in four sky regions, which should be a relatively simple case for the extraction to deal with, in order to focus on the differences between parametrizations.

We first note that the \textsc{ares} model plotted in Fig.~\ref{fig:comparison} has the typical features we expect: a maximum at $\sim 47\ \mathrm{MHz}$ corresponding to the onset of \lya\ coupling, a minimum at $\sim 71\ \mathrm{MHz}$ where the IGM starts heating, and a broad maximum at $\sim 111\ \mathrm{MHz}$ where ionization starts to become important. We will refer to these features as turning points B, C, and D, respectively (see Table \ref{tab:TP_desc}). If we interpolate between these maxima and minima using a cubic spline (the turning points parametrization), we produce the magenta dot--dashed line. Despite enforcing the correct parameter values, the shape of the curve under this popular parametrization scheme does not closely match the curve produced by the physical model. This might raise concerns that a turning points model will not perform well in the extraction of the signal from synthetic data, and these concerns are not dispelled by the solid blue line, which is the result of attempting this fitting. 

The overall difference in normalization between the black and blue lines comes about because a constant offset in the signal can be almost perfectly absorbed by the foregrounds, and so is very difficult to determine from the data. This can lead to an unphysical signal, which we can attempt to solve by imposing stricter priors, but which is not prevented by the parametrization itself. Even if we correct this offset in normalization by hand, however, we see that the shape of the recovered signal does not closely follow the shape of the input signal. This is shown by the dashed red line, for which we artificially add an offset to the recovered signal to ensure that the temperature of the minimum of the absorption trough matches the input. We can see that even the frequencies of turning points B and D are poorly recovered: B is at too high a frequency, and D is too low. A possible reason for this can be seen by comparison with the magenta dot--dashed line, which goes through the correct turning points. We see that near the minimum of the signal, the offset recovered curve captures the shape of the input signal much better. It seems, then, that by matching the shape of the signal between the turning points, we recover worse values for the positions of the turning points themselves. 

Finally, we examine a fit using the tanh model, shown with the dotted cyan line. This captures the overall shape of the signal much better than the turning points model, though there seems to be an offset at high frequency which we examine later. In fact, if we look at the position of the turning points recovered from the tanh model, they match those of the input signal better than the turning points model even though they are not explicit parameters of the tanh model. Moreover, since the tanh functions in this parametrization describe properties of the IGM, which are then translated into a \dtbbar, we can guarantee that the signal is physically plausible much more readily than for the turning points parametrization. The tanh model is, however, much faster to evaluate than a full \textsc{ares} run, and so it is a promising parametrization for which to test our fitting in the rest of this paper.

\section{Description of the recovery process} 
\label{sec:description}

Given a synthetic data set, i.e.\ a small number of independent, noisy spectra (from pointing towards different areas of the sky) including a 21-cm signal, foregrounds, instrumental response and noise, generated according to the procedure outlined in Section~\ref{sec:dareprob}, we wish to fit some parameters describing the different contributions to the data. In this paper we wish to concentrate our attention on the difference between parametrizations of the 21-cm signal, and so we fix the instrument model to be the same as the one used to generate the data set, and fit the parameters of a signal model (not necessarily the same one used to generate the data) and a straightforward foreground model. The global 21-cm signal is uniform over the sky at the angular resolutions and noise levels we consider here \citep{bittner2011}, so the parameters of the signal are the same in each sky area. The foreground parameters are allowed to differ, however.

We have updated our fitting routines from the ones used by \citet{harker2012} for an earlier version of the {\it DARE} pipeline. Fitting moderately large numbers of sky regions, more complex foregrounds, more computationally expensive signal models, and potentially parameters of an instrument model, requires us to be able to explore a complicated parameter space with perhaps a few dozen dimensions. For this reason, we now use the \textsc{emcee} package \citep{emcee2013}, which implements the affine-invariant Markov Chain Monte Carlo sampler of \citet{goodman2010}. This yields samples from the posterior probability distribution of the parameters of interest, given the data. In computing the likelihood, we assume all the frequency channels in all sky regions are independent, i.e.\ the probability density for obtaining the value $T^i_\mathrm{meas}(\nu_j)$, where $i$ indexes the sky region, for a vector of parameters $\boldsymbol{\theta}$, is
\begin{equation}
p_{ij} = \frac{1}{\sqrt{2\mathrm{\pi}\sigma_i^2(\nu_j|\boldsymbol{\theta})}}
\mathrm{e}^{-[T^i_\mathrm{meas}(\nu_j)-T^i_\mathrm{mod}(\nu_j|\boldsymbol{\theta})]^2/2\sigma_i^2(\nu_j|\boldsymbol{\theta})}\ ,
\label{eqn:pij}
\end{equation}
where $\sigma_i(\nu_j|\boldsymbol{\theta})$ is the rms noise in the
channel, computed from $T^i_\mathrm{mod}(\nu_j|\boldsymbol{\theta})$,
the bandwidth and the integration time using the radiometer equation, and the likelihood is just the product over all the channels,
\begin{equation}
L(\boldsymbol{T}_\mathrm{meas}|\boldsymbol{\theta}) = \prod_{i=1}^{\
N_\mathrm{sky\phantom{q}}}\prod_{j=1}^{n_\mathrm{freq}}p_{ij} \ .
\label{eqn:like}
\end{equation}
More generally, we could concatenate the individual spectra into a single `data vector', in which case our independence assumption implies a diagonal data covariance matrix, but this machinery is not necessary for the current work. In practice, we work with the log-likelihood, so the product in equation~\eqref{eqn:like} is computed as a sum. We adopt broad, Gaussian priors for the foreground parameters, which have little impact since the data generally constrain them quite well. For the signal parameters we generally adopt uniform priors; occasionally, these do come into play, for example in preventing the 21-cm signal parametrized by its turning points from becoming unphysical, but it is generally clear when this is the case, and we comment on it when relevant.

\subsection{Two-stage fitting}\label{subsec:twostage}

As alluded to in the previous section, we will also consider a scenario in which foreground removal has yielded the parameters of some 21-cm signal model, with errors, and we want to perform a second stage of inference about a different model. This situation might arise when the full parameter space has high dimension (for example, we wish to fit foregrounds in a large number of sky regions, perhaps including contributions from the Sun, Moon, etc., along with instrumental parameters) meaning that we can only perform the signal extraction using a signal model which can be generated rapidly. For example, we might infer the positions of the turning points, under that parametrization.

We might then wish to take those turning point constraints, and use them to infer something about the parameters of a model which is slower to evaluate, but has more physically meaningful parameters, for example the full \textsc{ares} signal model. We wish to test how much of the physical information in the data is retained in this two-step approach, since it may determine the requirements we place on our signal extraction pipeline. We also perform this second-stage fitting using \textsc{emcee}, but rather than the likelihood being computed as a sum over all the frequency channels in all the sky areas, it is computed assuming Gaussian errors on the parameters of the intermediate parametrization (either independent errors, or using the covariance matrix coming from the first-stage fitting).

\section{Recovering parameters from observations} 
\label{sec:recovery}

We start by comparing the turning points recovered by a direct fit of the `turning points' parametrization to an \textsc{ares} model with the turning points recovered by fitting the tanh parametrization (where in the latter case, the turning points are the extrema of the reconstructed signal).

\begin{figure*}
\includegraphics[width=\textwidth,clip=true]{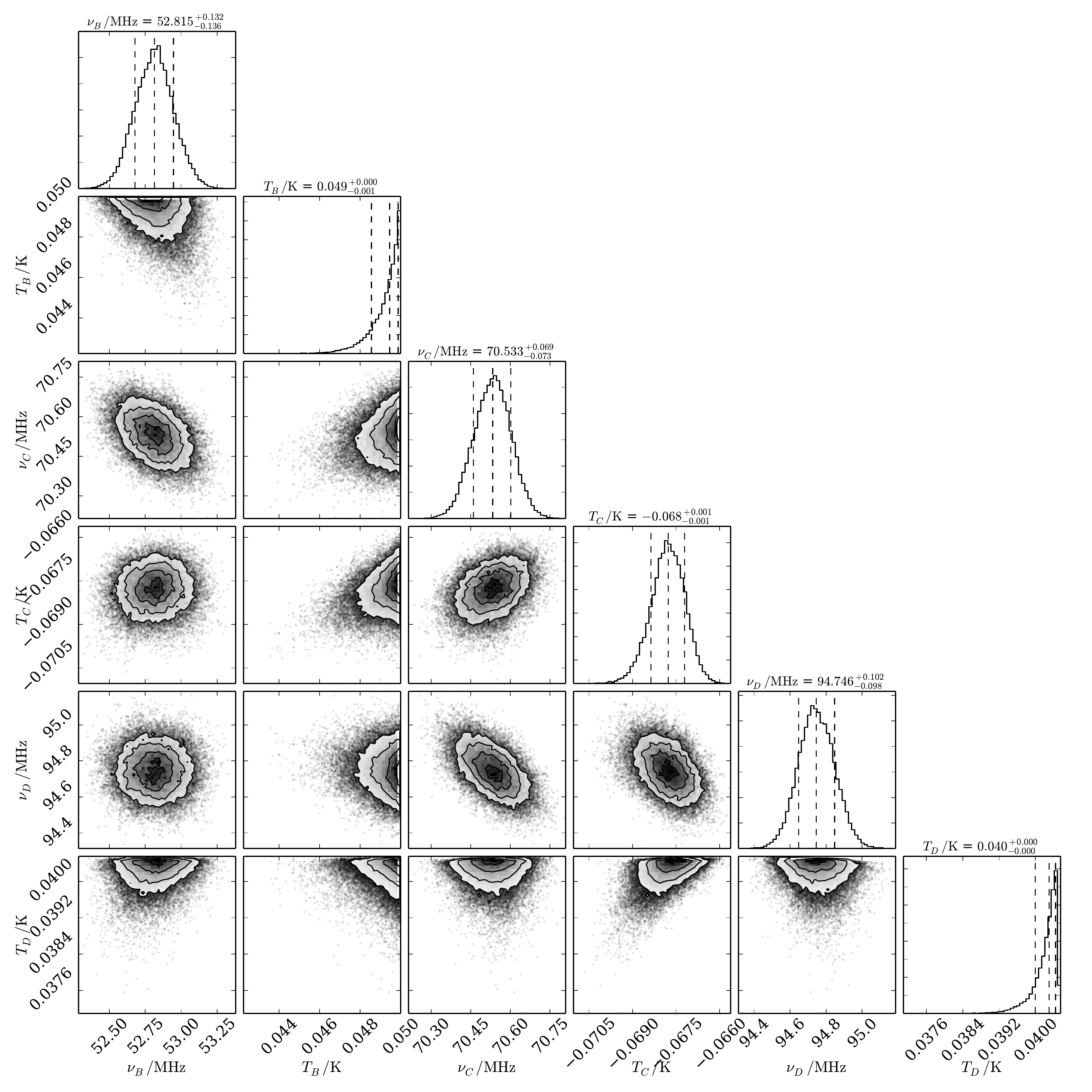}
    \caption{Turning points inferred from a direct fit of the `turning points' parametrization to the reference \textsc{ares} model, for an idealized antenna that has 85 per cent sensitivity across the whole band between 35 and 120 MHz, and for a total integration time of 1000~h spread across four sky regions. That is, the input signal is the solid, black curve of Fig.~\ref{fig:comparison}, while the fitted signal is the blue, solid curve of Fig.~\ref{fig:comparison}. This plot shows the marginalized 1D and 2D posterior probability distributions of the signal parameters, where $\nu_i$ and $T_i$ are the frequency and differential brightness temperature respectively of turning point $i$, where $i$ is B, C or D. $0.5\sigma$, $1\sigma$, $1.5\sigma$ and $2\sigma$ contours are shown with the solid lines and shading in each 2D panel, while individual samples of the posterior outside the $2\sigma$ contour are shown with individual dots. In the 1D histograms down the diagonal of the plot, the title above the panel gives the median value of the parameter, with errors calculated using the 16th and 84th percentiles of the distribution, while the vertical dashed lines in each panel show the 16th, 50th and 84th percentiles visually. For this fit, in no case does the `true' position of the turning point lie within the scale of the plot, even though the formal statistical errors on the positions of the turning points are relatively small. The distributions of $T_B$ and $T_D$ run up against the edge of the prior.}\label{fig:tpdirect_4_1000}
\end{figure*}

The direct fit of the turning points model, for the same case as for Fig.~\ref{fig:comparison}, is shown in Fig.~\ref{fig:tpdirect_4_1000}, which is a `corner plot' made using Foreman-Mackey's \textsc{triangle\_plot} package\footnote{\url{https://github.com/dfm/triangle.py}}.
As noted in the discussion of Fig.~\ref{fig:comparison}, the turning point constraints are biased, and all the `true' values of the turning point parameters lie outside the axis ranges of the panels in Fig.~\ref{fig:tpdirect_4_1000}. The statistical errors are therefore misleading, even though the instrument model is perfect and fixed, and the foreground model is able (by construction) to perfectly model the input foregrounds. None the less, some of the correlations we can see in the 2D posteriors, such as the positive correlation between the frequency and differential brightness temperature of turning point C, persist in other cases. For example, we see this correlation when the input model (as well as the fit mode) uses the `turning points' parametrization, or when the turning points are inferred from a tanh fit to the data.

\begin{figure*}
\includegraphics[width=\textwidth,clip=true]{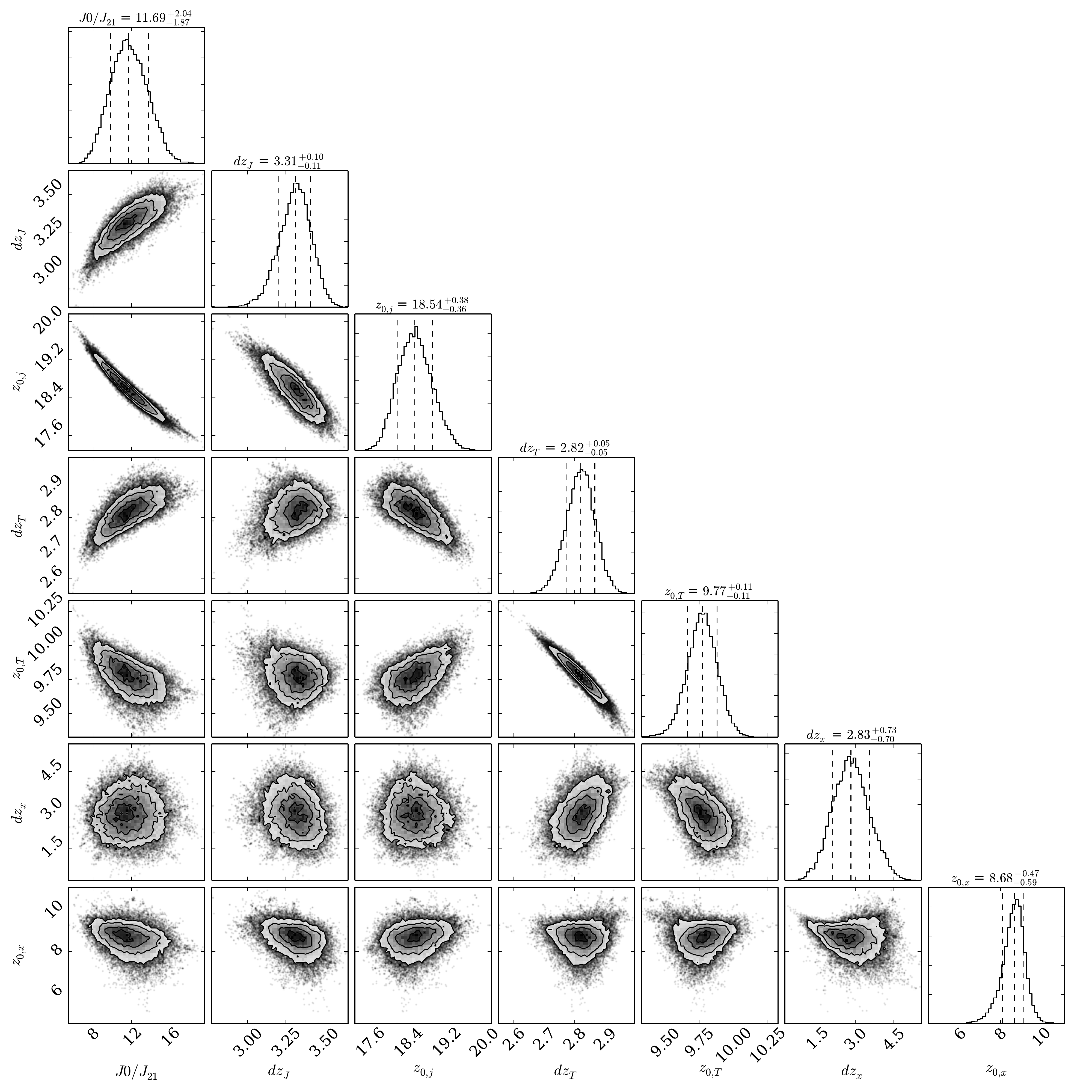}
    \caption{Parameters of the tanh model inferred from a fit to the reference \textsc{ares} model, for the same experimental setup as Figs.~\ref{fig:comparison} and \ref{fig:tpdirect_4_1000}. In order from left to right, the parameters are the normalization ($A_\mathrm{ref}$, see Equation~\ref{eqn:tanhdef}) of the \lya\ flux tanh function in units of $10^{-21}\ \mathrm{erg}\ \mathrm{s}^{-1}\ \mathrm{cm}^{-2}\ \mathrm{Hz}^{-1}\ \mathrm{sr}^{-1}$, the redshift interval ($\Delta z$) and the central redshift ($z_0$) over which the \lya\ background turns on, $\Delta z$ and $z_0$ for the X-ray heating, and $\Delta z$ and $z_0$ for the ionization step. The ionization step is fixed to a height of unity, since it represents a fraction, and the temperature step height is fixed to $1000\ \mathrm{K}$, since in practice the signal becomes saturated at low redshift, so the precise height of the step is not important.}\label{fig:tanhdirect_4_1000}
\end{figure*}

Similarly, Fig.~\ref{fig:tanhdirect_4_1000} shows the parameters of the tanh fit corresponding to the cyan dotted line in Fig.~\ref{fig:comparison}. 
Although there are some strong degeneracies, for example between the normalization and central redshift of the \lya\ history, and between the central redshift and width of the temperature step, which may suggest that a lower-dimensional parametrization could work, the constraints seem plausible and physical. Constraints on the reionization midpoint and duration are comparable to numbers quoted in the literature, but should be interpreted with caution as the bandpass used in the fit is truncated at 120 MHz, i.e., $z \approx 10.8$. As a result, these constraints would likely change if one imposed $z \lesssim 10$ prior information from the late stages of reionization. However, such constraints are still meaningful: we will focus on how these parameters are related to the volume filling factor of \ion{H}{ii} regions, $Q_{\ion{H}{ii}}$, and the volume-averaged ionization rate, $\Gamma_{\ion{H}{i}}$, momentarily.

\begin{figure*}
\includegraphics[width=\textwidth,clip=true]{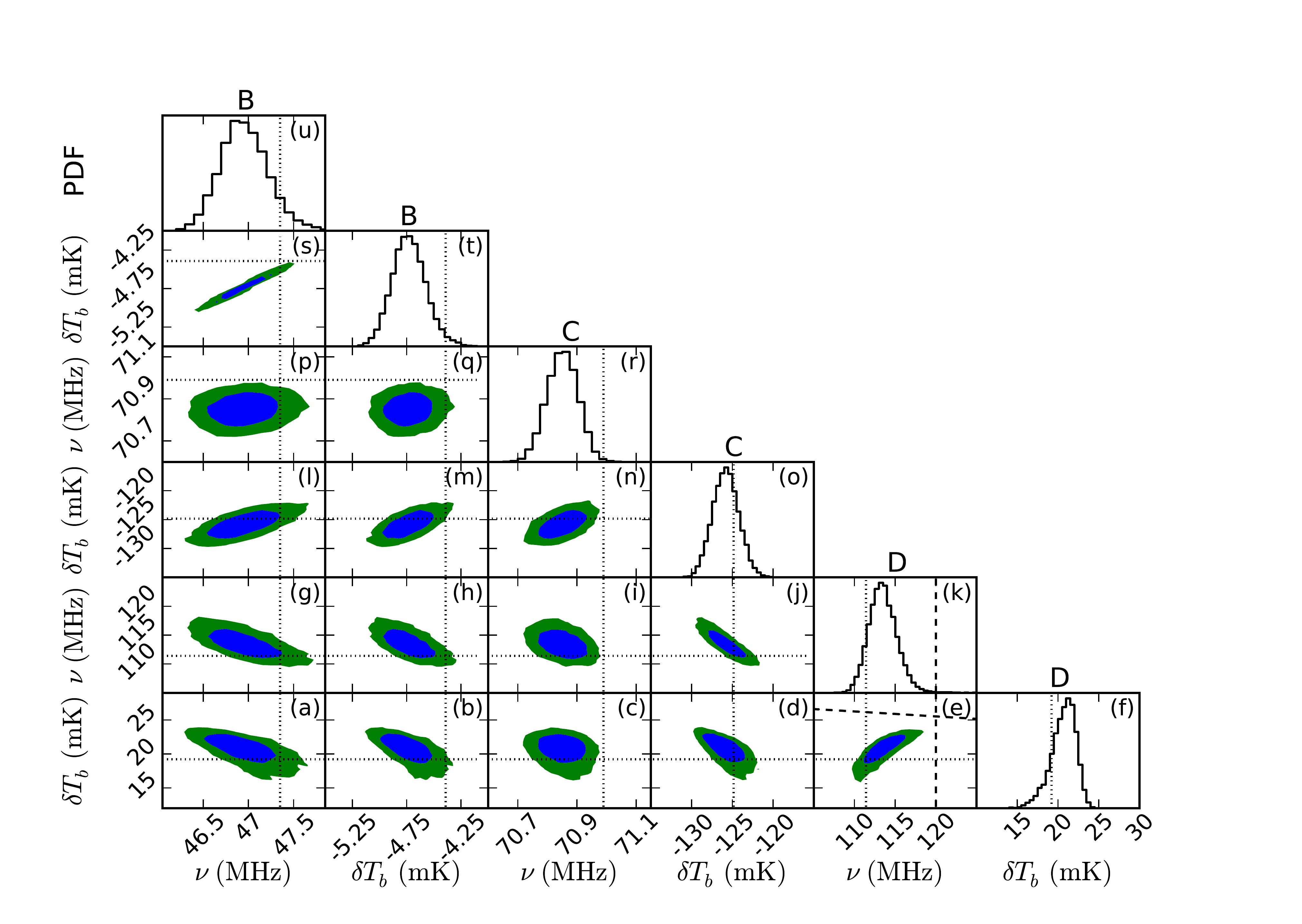}
    \caption{We show how the constraints on the \dtbbar\ history implied by Fig.~\ref{fig:tanhdirect_4_1000} translate into constraints on the turning points. In each panel, the dotted lines show the input parameter value. The dashed vertical line in panel (e) shows the upper end of the frequency range, while the nearly horizontal line shows the path the signal would follow in a hot, completely neutral Universe for which the emission signal saturates. The dark blue and green contours show $1\sigma$ and $2\sigma$ credible regions.}\label{fig:tpfromtanh_4_1000}
\end{figure*}

We see how these translate into constraints on the turning points in Fig.~\ref{fig:tpfromtanh_4_1000}.
The statistical errors are similar to those for the direct spline fit through the turning points (Fig.~\ref{fig:tpdirect_4_1000}), but the best-fitting values are clearly much closer to the true values, i.e.\ the bias is substantially reduced. Some biases remain, however, and are a consistent feature in all of our calculations, regardless of the integration time or number of sky regions. We have checked that despite the small residual bias, the fitting correctly captures the variation in turning point position with changes in \textsc{ares} parameters. For example, as the efficiency of X-ray production changes, the position of turning point C (in frequency and temperature) also changes, and the fitted values accurately reflect this.

We should expect biases in the turning point constraints to propagate to constraints on the IGM properties, so we will next investigate how to mitigate these biases. First, we test whether the constraints might be affected by the frequency range used for the fitting and the complexity of the foreground model. We can see from Fig.~\ref{fig:comparison} that there is a mismatch between the input model and the recovered tanh-fit at the highest frequencies, where it seems to lie below the input signal. This is not unique to the specific realization fitted in Fig.~\ref{fig:comparison}, and may simply be due to a degeneracy with the foreground, which is more difficult to overcome at high frequencies where the signal is smoothest. Biases at high frequencies, depending on the parametrization, could propagate to lower frequencies. A more complex foreground model may help by absorbing the smooth divergence between the tanh model and the input curve at high frequencies. Usually, this sort of degeneracy is a disadvantage since it weakens the constraints on the signal, but we aim to test whether, by allowing a wider range of histories at low redshift, a more complex model might avoid biases at higher redshift, perhaps at the expense of increased errors.

\begin{figure*}
\includegraphics[width=\textwidth,clip=true]{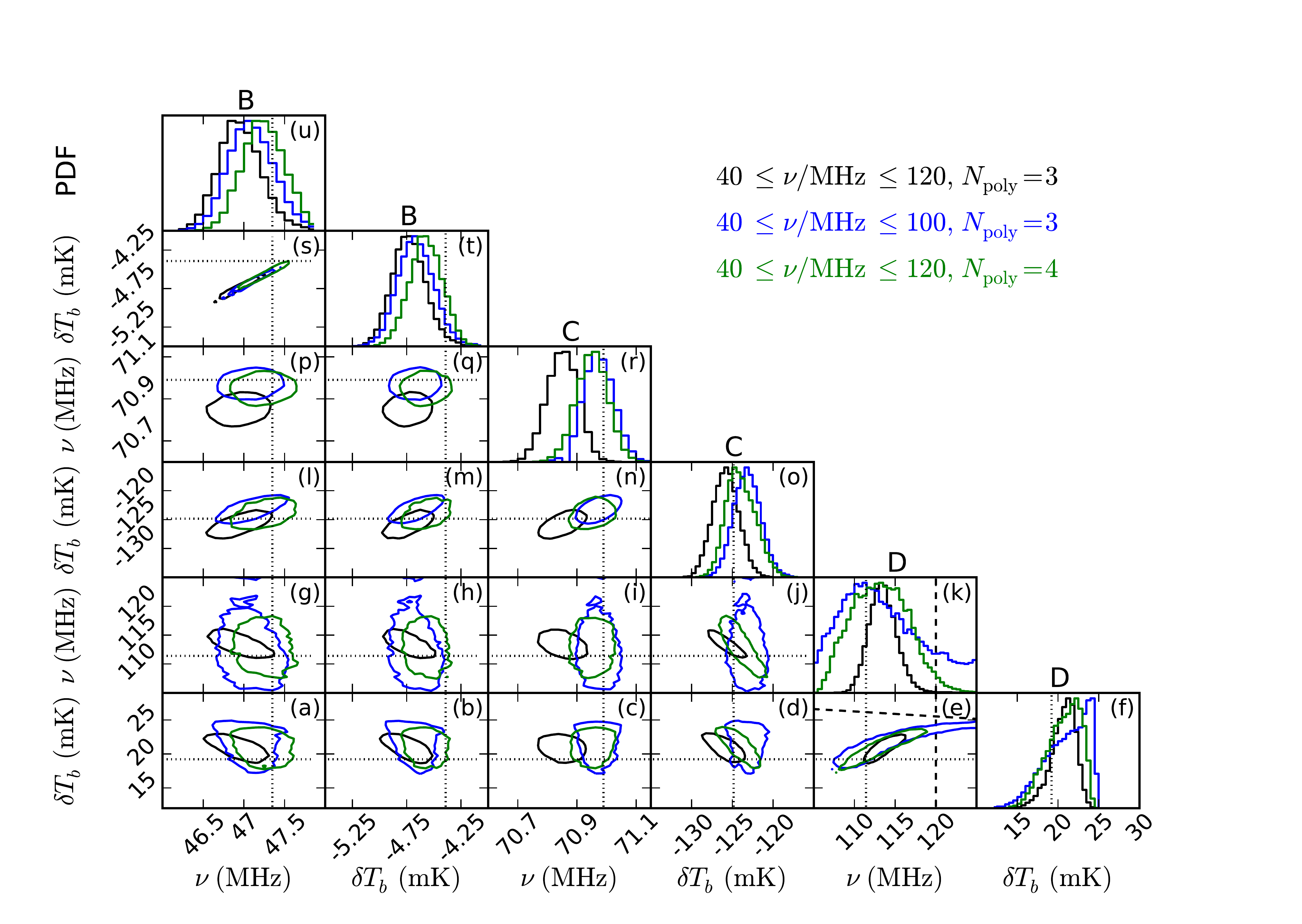}
    \caption{A comparison of the 68 per cent credible regions for constraints on the positions of the turning points for three different data sets: the same as that used in Fig.~\ref{fig:tpfromtanh_4_1000} (black); a data set where the upper limit of the frequency range is $100\ \mathrm{MHz}$ instead of $120\ \mathrm{MHz}$ (blue); and a data set where the foregrounds are fourth-order polynomials in $\log\nu$--$\log T$ rather than third-order (green).}\label{fig:std_100MHz_fg4}
\end{figure*}

The turning point constraints for the same model as in the earlier figures, for a data set truncated at $100\ \mathrm{MHz}$, and for a data set using the full frequency range but a fourth-order rather than a third-order polynomial model for the foregrounds, are shown in Fig.~\ref{fig:std_100MHz_fg4}.
As one would expect, truncating the frequency range significantly weakens the constraints on turning point D. It does, however, reduce the bias on the frequency of turning point C, for which the constraints were only marginally consistent with the true value in the standard case. Therefore it does seem that discarding frequencies where the parametrization is unable to capture the shape of the signal helps with recovery. Perhaps surprisingly, increasing the complexity of the foreground model has a similar effect, increasing the errors on the high-frequency turning point but reducing the bias on turning point C. The extra foreground parameters act as extra nuisance parameters which absorb the difference between the tanh model and the data, and which are marginalized over to produce an unbiased constraint.

\begin{figure} \includegraphics[width=\columnwidth,clip=true]{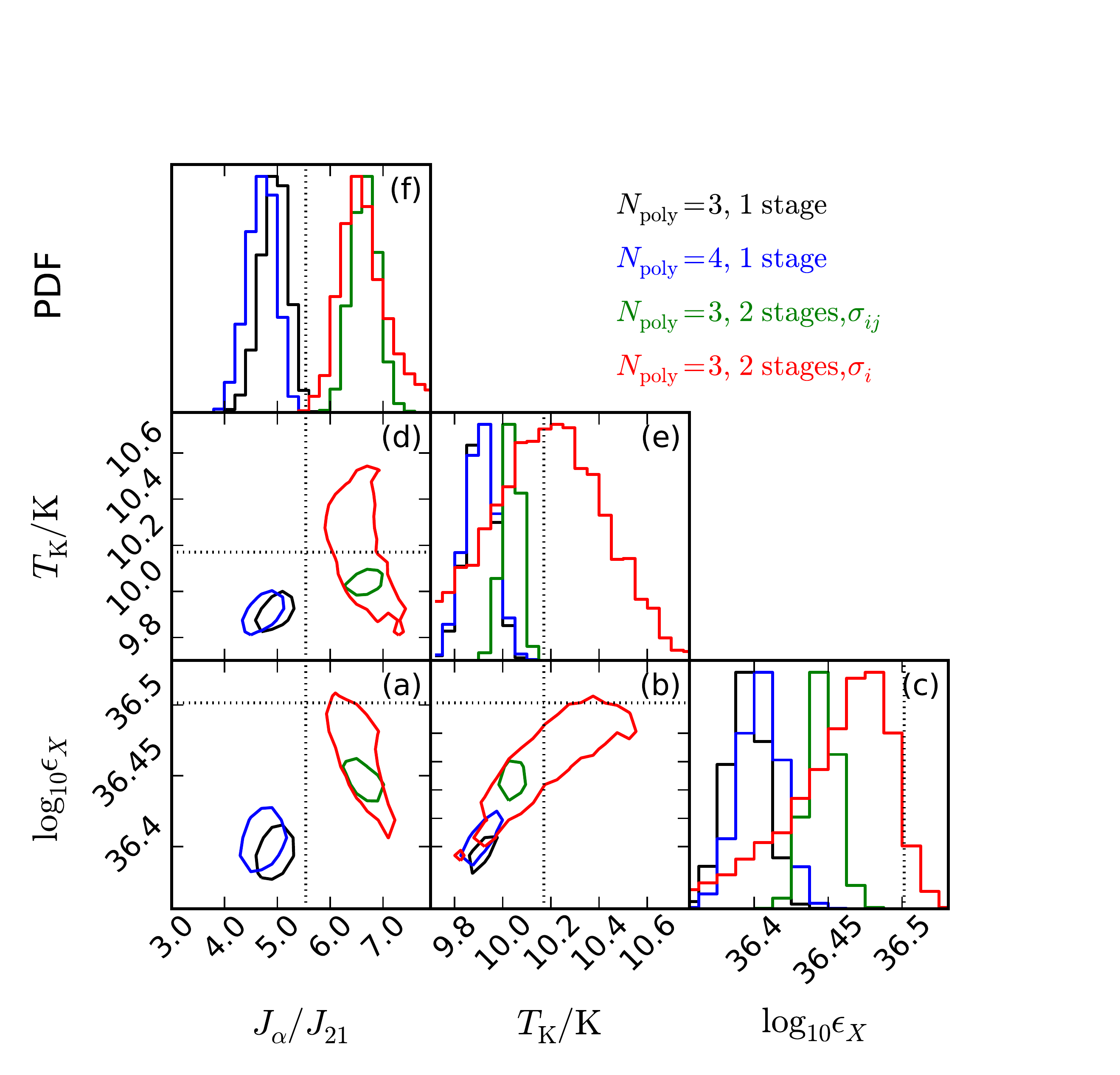}
  \caption{The constraints on the IGM parameters for which the constraining power comes mainly from turning point C. These parameters are the kinetic temperature of the gas (colder than the CMB at this point), the heating rate density ($\mathrm{erg} \ \mathrm{s}^{-1}$ per unit comoving $\mathrm{Mpc}^{3}$), and the \lya\ flux (in units of $J_{21} = 10^{-21} \mathrm{erg} \ \mathrm{s}^{-1} \ \mathrm{cm}^{-2} \ \mathrm{Hz}^{-1} \ \mathrm{sr}^{-1}$). The data set assumes 1000~h of data split between four sky regions, though the results are qualitatively similar for fewer sky regions and for shorter integrations. The black lines show constraints (1D posterior distributions and 1-$\sigma$ contours) coming directly from the tanh fit to the data set, while the blue lines show the results obtained with a more complex foreground model. The green lines assume that only the positions of the turning points and the covariances between the turning points are known, whereas the red lines assume that only the turning point positions are known (nothing about the shape of the signal in between) and that the errors on the turning points are independent and Gaussian, with the positions and the size of the errors coming from the tanh fit. Dotted vertical and horizontal lines show the true values.}
  \label{fig:twostageC}
\end{figure}

\begin{figure*} \includegraphics[width=\textwidth,clip=true]{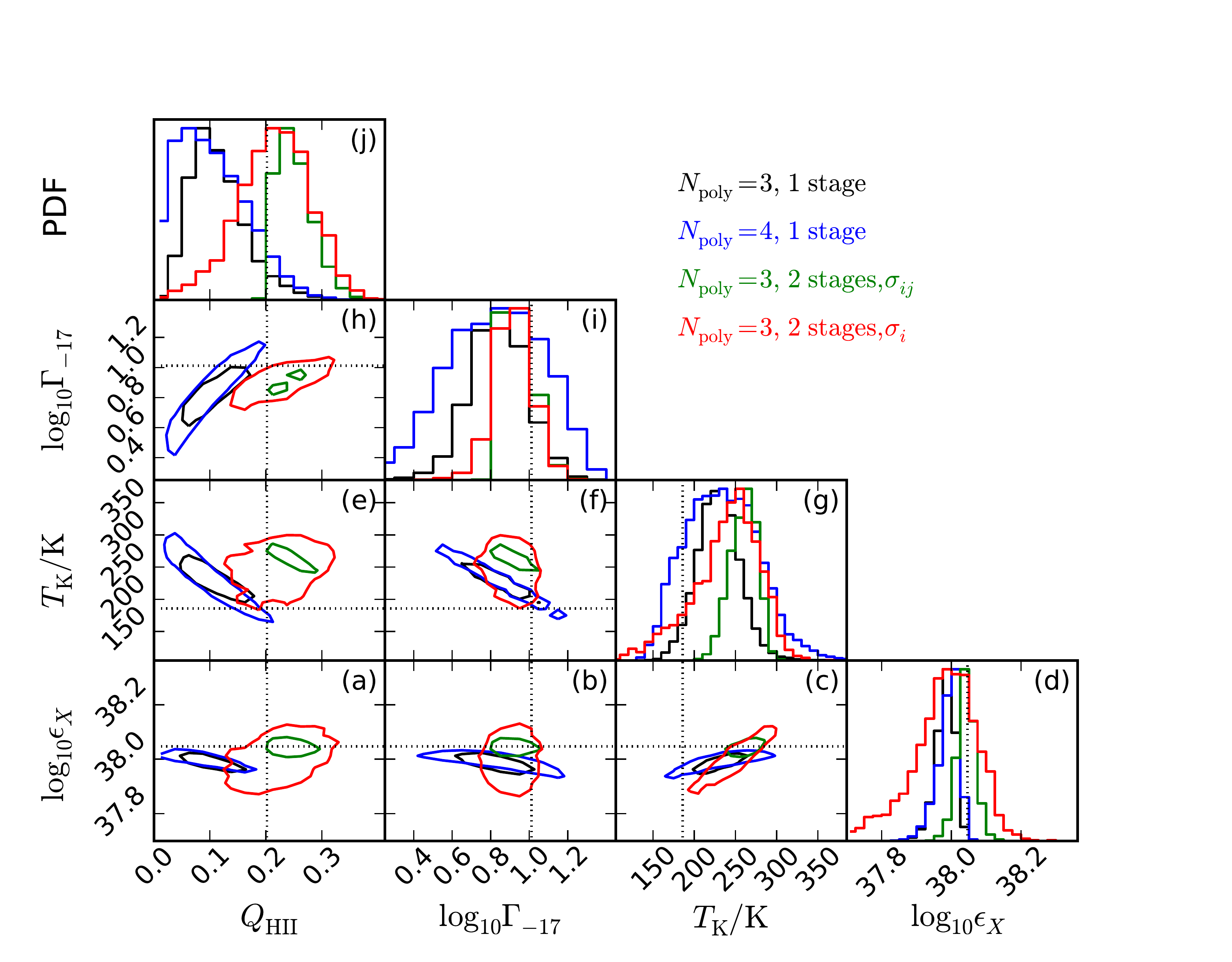}
  \caption{The constraints on the IGM parameters for which the constraining power comes mainly from turning point D. These parameters are the kinetic temperature of the gas (hotter than the CMB at this point), the heating rate density, the volume filling factor of \ion{H}{ii} regions, $Q_{\ion{H}{i}}$, and the volume-averaged ionization rate, $\Gamma_{\ion{H}{i}}$ (in units of $10^{-17} \ \mathrm{s}^{-1}$). The colours have the same meaning as those shown in Fig.~\ref{fig:twostageC}.}
  \label{fig:twostageD}
\end{figure*}

A similar effect can be seen at work if we examine one- and two-stage fits for the IGM parameters. By a one-stage fit, we mean that the properties of the IGM are taken directly from the tanh parametrization. By a two-stage fit, as discussed in \S\ref{subsec:twostage}, we mean that, first, a tanh fit is used to infer the positions of the turning points and then, secondly, the constraints on the turning points are used to infer IGM parameters. We look at two flavours of the second-stage fit in Figs~\ref{fig:twostageC} and \ref{fig:twostageD}.
In the first, we simply take the errors on the frequency and temperature of each turning point to be independent and Gaussian. In the second, we still assume the errors are Gaussian, but use the covariance matrix for the turning point parameters obtained from the posterior samples of the tanh fit.

If we only use the turning points to constrain the IGM parameters, and do not make use of any other shape information in the signal, this naturally increases the errors, as can be seen in the larger contours and broader 1D distributions for the two-stage fits. This is especially so if we disregard the correlation structure and treat each turning point parameter independently (red lines). The two-stage fits do, however, reduce the bias on the inference of the IGM properties. The black contours are inconsistent with the true values, while the red and green contours from the two-stage fits enclose the true value. Our interpretation of this is that although we have discarded shape information, retaining only the turning point positions, this shape information was unreliable, and biased our constraints. The turning points encode robust information about the signal, even when they are inferred from a parametrization (the tanh model) which does not explicitly include their positions as parameters. This highlights the importance either of finding a parametrization which is flexible enough to be able to capture the true shape of the signal, or finding robust quantities which yield reliable information even if the parametrization is imperfect. Of course, for a real experiment we do not know the shape of the signal in advance, though we may be able to choose between different parametrizations using e.g.\ the Bayesian evidence.

\begin{figure*} \includegraphics[width=\textwidth,clip=true]{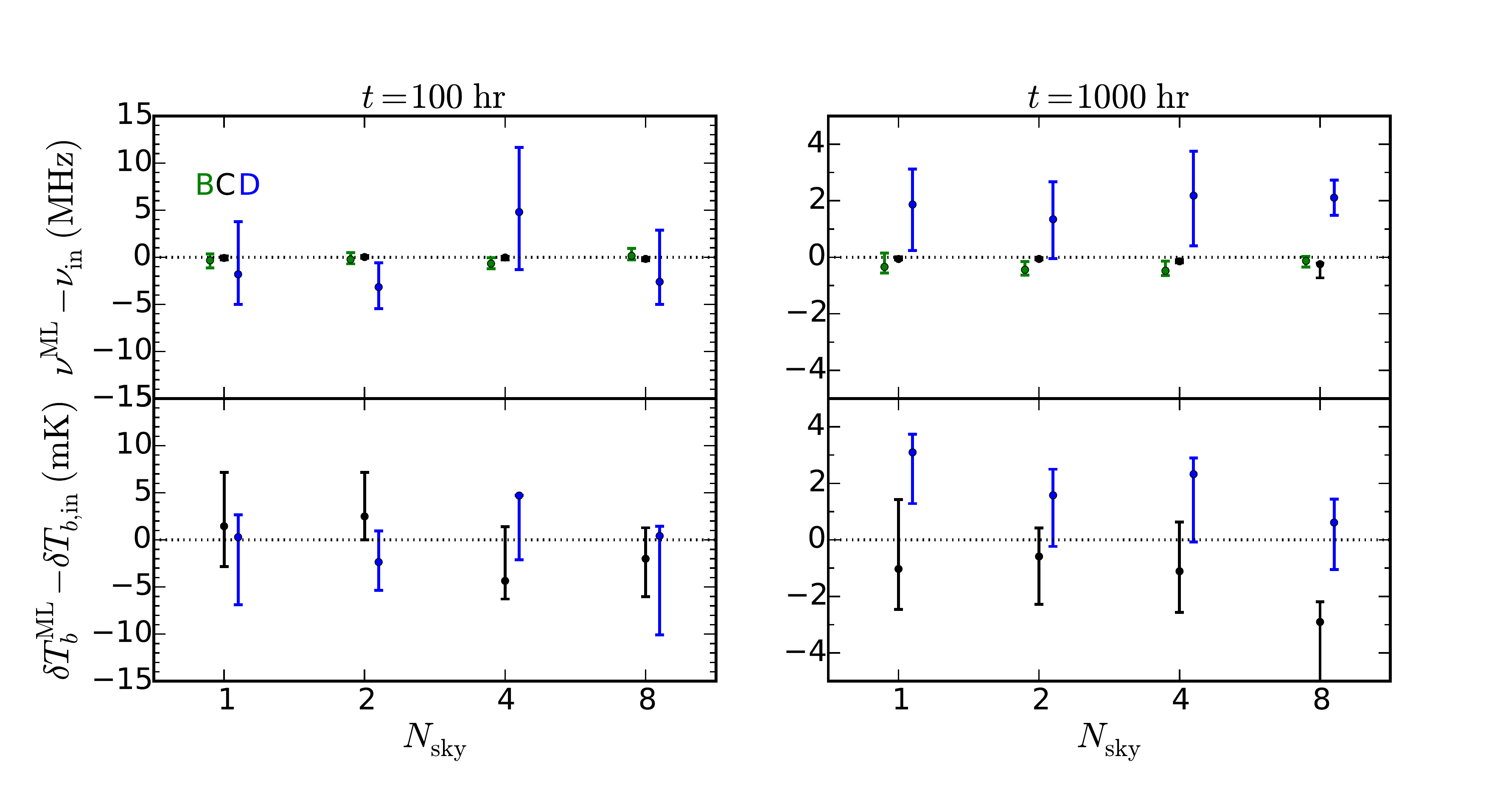}
  \caption{Constraints on the turning point positions as a function of the number of sky regions and integration time. Green, black and blue points correspond to constraints on turning points B, C and D, respectively, and are slightly offset in the $x$-direction for clarity. The top row shows errors in the frequency of the turning points, relative to their input values, while the bottom row shows errors in the amplitude of each turning point. Note that we do not show the errors on the temperature of turning point B, since these are too large to strongly constrain models, and would stretch the scale of the figure. All error-bars shown are 68 per cent credible intervals. Note that the $y$ range for the panels on the right has been zoomed in.}
  \label{fig:Nsky_tint}
\end{figure*}
In Fig.~\ref{fig:Nsky_tint}, we distil the results of a calculation suite in which the total integration time, $t_\mathrm{int}$, and the number of independent sky regions, $N_\mathrm{sky}$, is varied, showing constraints on the turning points as a function of $t_\mathrm{int}$ and $N_\mathrm{sky}$. Increasing the integration time by a factor of 10 has a much more substantial effect than increasing the number of sky regions, though subtle $\sim 1\sigma$-level biases, particularly in the frequency and brightness temperature of turning point D, are persistent even with 1000~h integration. Increasing $N_\mathrm{sky}$ from 1 to 2 has a greater impact than any further increases, though this conclusion may change if we must also fit parameters of the instrument, since we will then have constraints from a greater variety of signals being passed through the instrumental response. The results for $t_\mathrm{int}=1000\ \mathrm{h}$ and $N_\mathrm{sky}=8$ are perhaps surprisingly poor, but this reflects the fact that by using the entire sky, we end up including the parts with the most intense foregrounds, in the direction of the Galactic Centre. Since this paper's focus is on signal parametrizations, an investigation of the tradeoff between the number of independent sky regions and the intensity of the foregrounds (and how this affects constraints on the instrumental response) is beyond its scope.

\section{Discussion} 
\label{sec:disc}

While these calculations can provide some indication of the most useful or flexible parametrizations to use for signal extraction for observational data, it would be more useful to have an indication of which is the best model to use from the data themselves. This may be supplied by a computation of the Bayesian evidence, but although \textsc{emcee} provides in principle for the computation of the appropriate marginal likelihood using its parallel tempering mode, we have found that to be impractical. We have used nested sampling \citep{skilling2004} to perform model selection for an idealized and simplified version of the problem we study here \citep{harker2015}, but we found that \textsc{multinest} \citep{feroz2009} did not scale well enough to apply to the problems in the current paper, and an alternative such as \textsc{polychord} \citep{handley2015} may be required.

In principle, the best parametrization might also depend on the properties of the instrument, since instrumental parameters might be more degenerate with some signal parametrizations than with others. If the instrumental response is known, but is different from the response we have assumed here (a constant 85 per cent efficiency across the band), our conclusions are unaffected, though the errors on the parameters can change owing to the change in sensitivity. We have checked this using a modelled instrumental response for \emph{DARE}. If we must fit instrumental parameters, then the conclusions may depend in detail on the properties of the instrument, and so are beyond the scope of this paper, which attempts a more general discussion.

\section{Conclusions}
\label{sec:conc}

We have described a `tanh' model for the global 21-cm signal between the end of the dark ages and the start of the epoch of reionization, which employs simple parametric forms for the \lya\ background, IGM temperature and reionization histories, and which matches the shape of physical models much better than the `turning points' model used in previous work. The tanh model also helps pin down the overall normalization of the signal, and thus the position of its turning points, despite the fact that the turning points are not explicit parameters of the model. This is largely because the tanh model has stronger theoretical priors, e.g., the `dark ages' feature is confined to a narrow `track' at $\nu \lesssim 50$ MHz, and the signal cannot become `super-saturated' at late times (low redshifts). Moreover, by describing IGM properties explicitly, it opens the door to including other constraints on the reionization history (e.g.\ from the CMB) in our likelihood function. It does all this while being several orders of magnitude faster to compute than a full physical model of the 21-cm signal, allowing us to explore the large parameter spaces which are required if we are to simultaneously fit parameters of the signal, foregrounds and instrument. We can none the less take the parametric fits and use them to constrain simple galaxy formation models, as shown by \citet{mirocha2015}.

We have found that integration time plays a larger role than the number of independent sky areas in the quality of signal recovery, though subtle biases persist in turning points constraints, particularly at the highest frequencies ($\nu \gtrsim 100$ MHz). This can be remediated by a more complex foreground model or by truncating the band at $100$ MHz, though the latter renders all constraints on the IGM at the lowest redshifts meaningless.

Even when the turning point constraints are unbiased relative to the input values, inferences of the properties of the IGM at the turning points can be biased. This is due to a subtle mismatch in shape between the tanh model and the \textsc{ares} physical model, which can be seen upon evaluation of the curvature at the turning points. In `two-stage fits', one can mitigate such effects by treating the errors on the turning points as independent Gaussians: while this is admittedly a more conservative estimate of the errors, it is a treatment which removes most knowledge of the detailed shape of the signal, keeping only the information which is more robust.

Finally, a direct `single-stage fit' using the parameters of the \textsc{ares} model might well be ideal, and would provide a useful point of comparison to our two-stage fits. For example, it would be interesting to see if biases in IGM properties would persist even if the signal were fit with the exact model used to generate it. We did not find this to be computationally feasible for this work, which highlights the need for future work to consider other samplers to explore our parameter space, and to tackle model selection as well as parameter estimation.

\section*{Acknowledgements}

GH acknowledges funding from the People Programme (Marie Curie Actions) of the European Union's Seventh Framework Programme (FP7/2007--2013) under REA grant agreement no.\ 327999. JM acknowledges support through the NASA Earth and Space Science Fellowship programme, grant number NNX14AN79H. JRP acknowledges support under ERC-2014-STG grant \#638743-FIRSTDAWN, FP7-PEOPLE-2012-CIG grant \#321933-21ALPHA, and STFC consolidated grant ST/K001051/1. This research has been supported by the NASA Ames Research Center via Cooperative Agreements NNA09DB30A and NNX15AD20A, and by NASA ATP grant NNX15AK80G. This work utilized the Janus supercomputer, which is supported by the National Science Foundation (award number CNS-0821794) and the University of Colorado Boulder. The Janus supercomputer is a joint effort of the University of Colorado Boulder, the University of Colorado Denver and the National Center for Atmospheric Research.

\bibliographystyle{mnras}
\bibliography{globalbib}






\bsp	
\label{lastpage}
 \end{document}